%% file: main.tex
\documentclass{article}

\usepackage{spconf, amsmath, graphicx}
\usepackage{times}
\usepackage{epsfig}
\usepackage{amssymb}
\usepackage{qiangstyle}
\usepackage{multirow}
\usepackage{enumitem}
\usepackage{subcaption}

\usepackage[ruled,vlined]{algorithm2e}

\title{Omni-sparsity DNN: Fast Sparsity Optimization for On-Device Streaming E2E ASR via Supernet}
\name{\parbox{0.95\linewidth}{\centering {Haichuan Yang$^*$\thanks{$^*$Equal Contribution}, Yuan Shangguan$^*$, Dilin Wang$^*$, Meng Li, Pierce Chuang, Xiaohui~Zhang,~Ganesh~Venkatesh, Ozlem Kalinli, Vikas Chandra}}}

\address{Meta AI}

\begin{document}

%
\maketitle
\input{icassp/abtract}


\input{icassp/intro}
\input{icassp/method}

\input{icassp/setup}

\input{icassp/exp}


\input{icassp/conclusion}

\bibliographystyle{IEEEbib}
\bibliography{asr}

\end{document}

%% file: icassp/abtract.tex
\begin{abstract}
%
From wearables to powerful smart devices, modern automatic speech recognition (ASR) models run on a variety of edge devices with different computational budgets. 
To navigate the Pareto front of model accuracy vs model size, researchers are trapped in a dilemma of optimizing model accuracy by training and fine-tuning models for each individual edge device while keeping the training GPU-hours tractable.
%
%
In this paper, we propose Omni-sparsity DNN, where a single neural network can be pruned to generate optimized model for a large range of model sizes.
We develop training strategies for Omni-sparsity DNN that allows it to find models along the Pareto front of word-error-rate (WER) vs model size while keeping the training GPU-hours to no more than that of training one singular model.
We demonstrate the Omni-sparsity DNN with streaming E2E ASR models. Our results show great saving on training time and resources with similar or better accuracy on LibriSpeech compared to individually pruned sparse models: 2\%-6.6\% better WER on Test-other.
\end{abstract}

%% file: icassp/intro.tex
\vspace{-2mm}
\section{Introduction}
\label{sec:intro}
\vspace{-2mm}

End-to-End (E2E) automatic speech recognition (ASR) models have gained popularity for deployment on edge devices~\cite{he2019streaming,Reviewondevice}. Neural network pruning is one of the key techniques to reduce E2E ASR model size while maintaining reasonable model accuracy.
However, ASR models usually live in a variety of edge devices -- from wearables with tiny RAM to powerful accelerator-enabled smart devices -- each with its own set of computational constraints. 
How to efficiently optimize the WERs of models on different devices without the burden of repeatedly training is an open challenge.

Recent works explored training multiple models with one framework, e.g.,  \cite{yu2019universally,yu2020bignas, cai2020onceforall, wang2020hat, nagaraja2021collaborative, mohtashami2021simultaneous}. 
The key idea is to encapsulate different models into a \emph{supernet}.
A supernet is a weight-sharing graph, wherein each model lives as a sub-network in the supernet.
The supernet training coordinates the updates of all sub-networks, and 
optimizes a single set of parameters that ensure all sub-networks simultaneously reach good performance at the end of the training.
Hence once the supernet is trained, one can run typical search algorithms, e.g., evolutionary search, to find the best models that satisfy the resource constraints of interest.
This search process is often efficient since there is no need of re-training and fine-tuning. 

%
%
%
Inspired by the success of the supernet, in this work,  we propose the construction of a supernet for E2E ASR, dubbed Omni-sparsity DNN, to efficiently explore the search space of sparsity in ASR models. By applying pruning masks (i.e. masks with zeroes) on each layer of the Omni-sparsity DNN weights, we can sample many sub-networks of different sizes, whose weights are shared and optimized jointly in the Omni-sparsity setup. 
%
The proposed Omni-sparsity DNN thus 
enjoys two key benefits that the supernet framework provides.
Firstly, it maintains a single set of supernet model weights, while generating sparse models for any target sparsity. The generated sparse models satisfy different device constraints along the word-error-rate (WER) vs model-size Pareto front -- they are on-par or better than individually optimized models. Secondly, Omni-sparsity supernet requires the similar amount of training time and resource of training one single sparse model.


Most related to our approach, the DSNN~\cite{wu2021dynamic} jointly optimizes a number of sparse models with pre-defined sparsity ratios by sharing their underlying model parameters. The key drawback of DSNN is that it trains a few candidate sub-networks with fixed model configurations. Not only does DSNN require large amount of training resources, when a new edge device with a distinct model-size constraint is presented to DSNN, it also needs to retrain the models.
The Omni-sparsity DNN, however, could directly generate an optimized model to fit this new device.

In Section~\ref{sec:training}, we describe the Omni-sparsity supernet training and the sub-network searching mechanisms. We develop three key strategies to guarantee the accuracy of Omni-sparsity DNN training under a tight training budget: 
1) an efficient in-batch sandwich sampling strategy to sample the sub-networks from the supernet such that all sub-networks will be sufficiently optimized;
2) a robust pruning criterion, called \emph{Adam-pruning}, to generate consistent pruning mask during the training, and consequently stabilizing the training procedure of Omni-sparsity DNN;
3) an adaptive dropout scheme that regularizes different sub-networks to different extents according to their model capacities.
%
We demonstrate the results of the Omni-sparsity optimization scheme with streaming recurrent neural network transducers (RNN-Ts) speech models (see Section~\ref{sec:exp}). With one training job, the Omni-sparsity DNN finds a family of sparse models, with 50\%- to 80\%-sparsity, that perform on-par with individually trained models on Librispeech test-clean set, and 2\%-6.6\% better in WER on test-other (see Section~\ref{sec:results}).

%% file: icassp/method.tex
\vspace{-2mm}
\section{Efficient Omni-Sparsity DNN Optimization}\label{sec:training}
\vspace{-2mm}

%
%
In this section, we introduce our Omni-sparsity DNN optimization method. Specifically, we train a dense supernet which can directly generate different sparse sub-networks by masking the model weights, and then use evolutionary search to find the best sparsity configurations for different model size budgets.
%
\begin{figure}[t]
\centering
\includegraphics[width=0.45\textwidth]{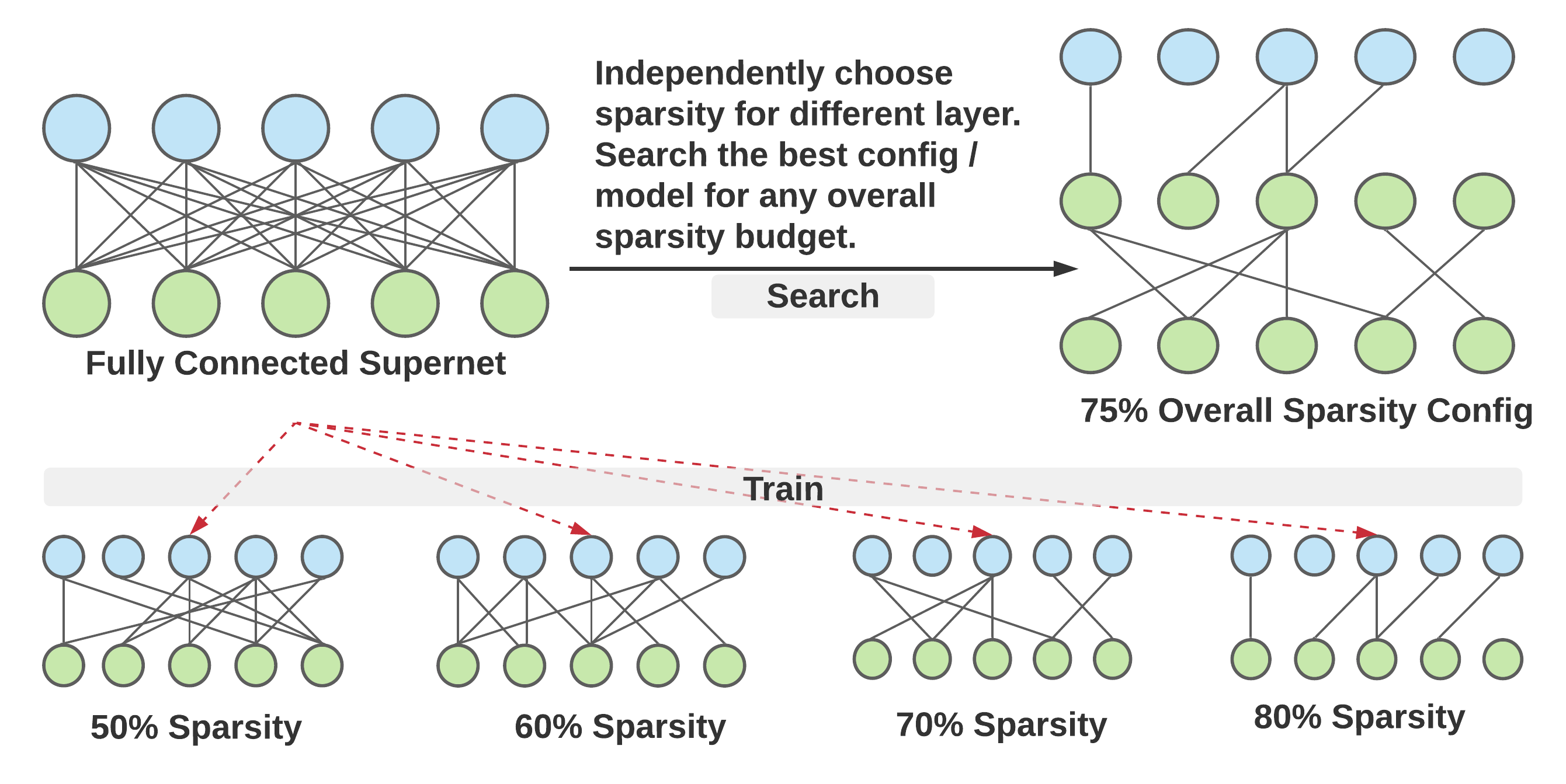}
\caption{During training, the supernet samples layerwise sparsity ratios, and applies the corresponding sparse masks to compute the gradients of the shared supernet weight. After training, we use an evolutionary search scheme to find the optimized models along the Pareto front of different model sizes.}
\label{fig:supernet}
\vspace{-3mm}
\end{figure}

\begin{figure*}[ht]
\centering
\setlength{\tabcolsep}{10pt}
\begin{tabular}{cc}
\centering
\raisebox{2.0em}{\rotatebox{90}{WER (test-other)}}  
\includegraphics[height=0.2\textwidth]{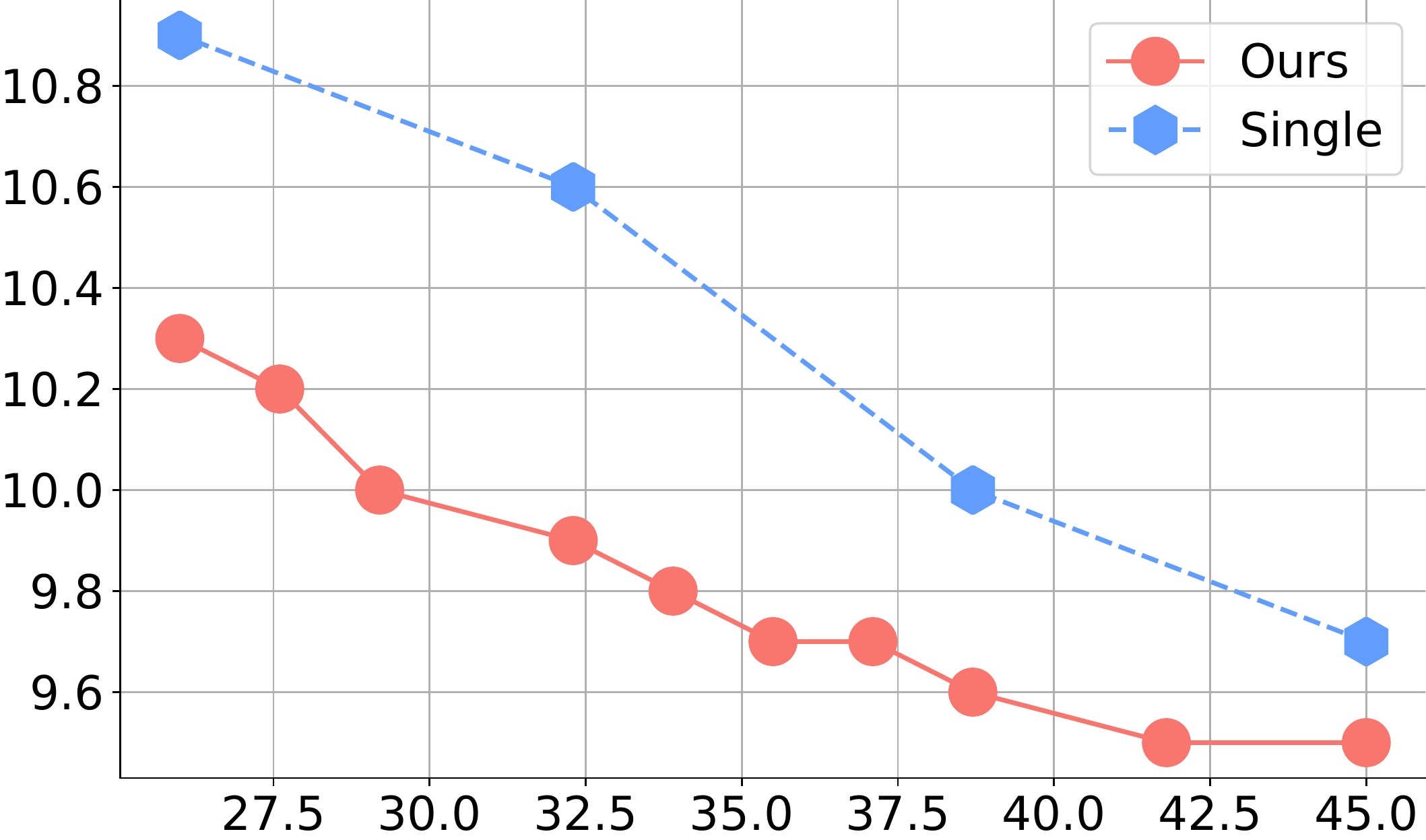} &
\raisebox{2.0em}{\rotatebox{90}{WER (test-other)}}  
\includegraphics[height=0.2\textwidth]{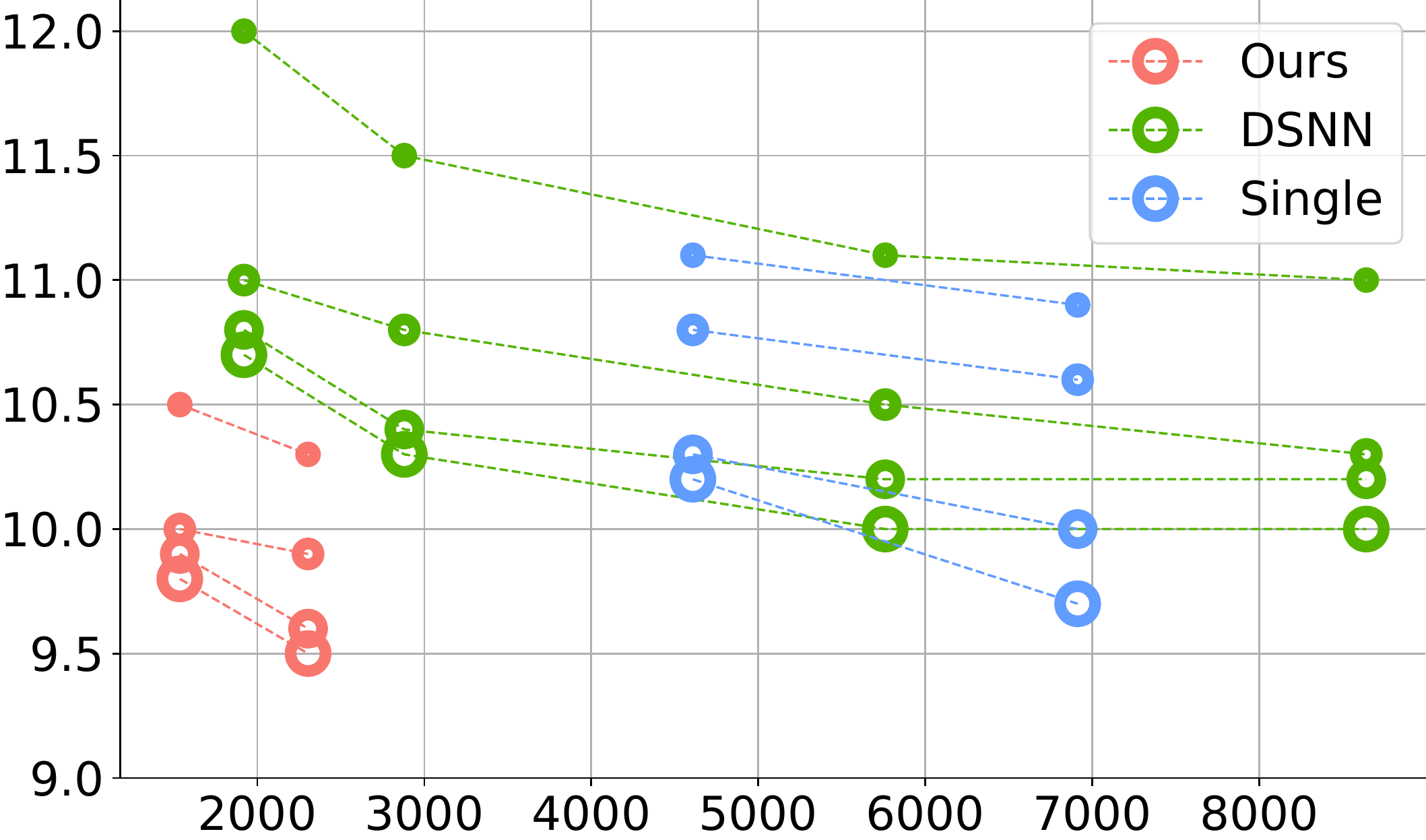} \\
(a) Model size (MB) & 
(b) Total training time (GPU hours) \\
\end{tabular}
\vspace{-2mm}
\caption{Training time v.s. WER for different sparse models. Marker size in (b) is proportional to the number of nonzero weights.}
\vspace{-2mm}
\label{fig:train_time}
\end{figure*}

\vspace{-2mm}
\subsection{Supernet-based layer-wise sparsity training}\label{sec:supernettraining}
\vspace{-2mm}

Let $\theta$ be the set of the model parameters, and let $L$ be the number of layers in the DNN. We define the layerwise sparsity ${\vv s}:=[s_1, s_2, ..., s_L] \in \mathcal{Q}$, 
where $\mathcal{Q}$ is a search space that contains all the possible sparsity configurations, e.g., $\{0.5, 0.6, 0.7\}^L$.
We denote $\theta_{{\vv s}}$ the corresponding weights parameters of the sparisfied sub-network by applying the pruning mask with sparsity $\vv s$. 

To train a supernet such that all its sub-networks simultaneously reach good performance can be formulated as the following optimization problem:
\begin{equation}
    \min_\theta \E_{{\vv s} \sim \mathcal{Q}}\bigg[ \E_{(x, y)\sim \D^{trn}}~\ell(y,x; \theta_{\vv s}) \bigg],
    \label{eq:train}
\end{equation}
where $\D^{trn}$ is the training data and $\ell(\cdot)$ represents the training loss, e.g., $\ell(y,x; \theta_{\vv s})=-\log p(y | x; \theta_{ {\vv s}})$, is the RNN-T transducer loss~\cite{graves2012sequence} computed with alignment restrictions~\cite{mahadeokar2021alignment}.



Eqn~\ref{eq:train} poses three main challenges to our model training and optimization:
1) Since we sample layerwise sparsity in the search space $\mathcal{Q}$ during training, we need to find ways to guarantee that all sub-networks can be sufficiently sampled and optimized within reasonable training budget.
2) to obtain highly accurate sub-networks, supernet training requires stable gradients; sampling new sparse masks at each training step induces instability in the supernet gradient.
3) the supernet contains both over-fitting (for the dense model) and under-fitting (for sparse models) tendencies, making regularization during training tricky. We therefore propose three training mechanisms to address these issues.\\
\textbf{1) Efficient in-batch sandwich sampling:}
Ideally, one would like to sample as many sub-networks as possible during training to ensure the convergence of all sub-networks in the supernet. 
Large sampling size, however, incurs large training costs.
Motivated by the sandwich sampling rule in the Slimmable networks ~\cite{yu2018slimmable}, at each training step, we only sample four sub-networks: the smallest, the largest and two random ones.
Meanwhile, to ensure the total training cost of the supernet is comparable to the cost of training one single network,
we limit each sub-network to only see a portion of the training batch at each step. 
More precisely, given an mini-batch $B$, 
we split the batch into four parts $\{B_1, \cdots, B_4\}$ with equal size, and train each of the four sub-networks on these parts separately.
This can be conveniently implemented with distributed data parallel,
such that each machine samples a different sub-networks but the batch gradients are aggregated from all machines. Additionally, we leverage knowledge distillation to accelerate the training of small sub-networks, similar to the recommended supernet training practice in the literature~\cite{yu2020bignas,wang2021alphanet}. 
Each sparse sub-network learns from the logits produced by the corresponding dense supernet. 
Note that the training cost will increase by 18\% in GPU-hours with in-place knowledge distillation. \\
\textbf{2) Robust sparse mask generation:} Each sub-network drawn from the supernet is pruned via a sparse mask following a predefined pruning criterion. In~\cite{shangguan2019optimizing}, the authors used a weight-magnitude based pruning criterion to decide which weights to zero out. In practice, using the gradient information can achieve better pruning results~\cite{lee2018snip}.
Consider a gradient-based criterion where we characterize the importance of a weight/connection, $w$, as:
\begin{equation}
|\ell(w) - \ell(0)| \approx |w| |\nabla_w \ell|.
\label{eq:adam_pruning}
\end{equation}
The gradient term in Eqn~\eqref{eq:adam_pruning} is often noisy; its variance is exacerbated by supernet sampling different sparsity configurations at each training step. 
To improve supernet performance, and at the same time stabilize the pruning criterion, we propose the use of
moving average of gradients to stabilize the pruning criterion.
Specifically, we replace $|\nabla_w \ell|$ with the square root of the moving average of the second order gradient moments in Adam. 
And we refer to our pruning algorithm as \emph{Adam-pruning}, which is still as efficient as the typical weight-magnitude based or gradient-based pruning approach, but it's more suitable for the supernet training due to dampening effect from the large momentum term on the gradients. \\
{\bf 3) Adaptive dropout:} 
Dropout is an important regularization technique to reduce over-fitting in E2E ASR models. Previously, an unchanged dropout value is applied to the layers of a network throughout the training process. The sub-networks in a supernet, however, over-fit to the training data to different degrees, and thus find it sub-optimal to adopt the same drop-out regularization during training. 
Intuitively, a large dropout for sparse models will likely cause under-fit; a small dropout for dense models might lead to over-fit. 
We propose adaptive dropout -- we regularize different sub-networks to different degrees according to the sub-network's modeling capacity.
To do so, we set the dropout rate based on the sparsity setting \emph{on the fly}. Specifically, for each layer with sparsity $s$, we set its dropout rate to $0.1 * (1.0 - s)$. Empirically, we find that adaptive dropout dramatically improves the WER of supernet-trained models. 

In-batch sandwich sampling, Adam-pruning and adaptive dropout allow us to train the supernet and its sub-networks effectively with a small training budget. We examine their efficiency and effectiveness in our ablation studies (Section~\ref{sec:ablation}).

\vspace{-2mm}
\subsection{Supernet-based Pareto Searching}
\vspace{-2mm}

After training the supernet, we expect all its sub-networks, with different sparsity configurations, well optimized. To find a set of sparse models with the best WER vs. run-time efficiency could then be solved with an evolutionary search.
For example, consider a set of model size constraints $\{\tau_1, \cdots, \tau_k\}$, finding the corresponding optimal sub-networks from the supernet that satisfy the constraints can be achieved as follows,
\begin{equation}
    \bigg\{ \min_{\vv s_i}~~\E_{(x, y)\sim \D^{val}}\bigg[ \ell(y,x; \theta_{\vv s_i}) \bigg],~~~ \text{s.t.}~ \mathcal{M}(\vv s_i) \le \tau_i \bigg\},
    \label{eq:evo_search}
\end{equation}
where $\mathcal{M}(s_i)$ denotes the model size of the sparse models with sparsity configuration $s_i$. $\D^{val}$ represents the validation dataset. 
We use loss on the validation set as the surrogate metric to rank the performance of different sub-networks.
For each evolutionary search iteration, we perform random mutate and cross-over for the layerwise sparsity configuration on the current Pareto front, and then compute their validation loss and update the new Pareto front.
Optimal sub-networks determined by the evolutionary search can be directly sampled from the supernet without the need of fine-tuning and retraining. 
The overall search cost is orders of magnitude lower than training: 50 GPU-hours to search and compute validation loss for 4000 networks on Librispeech.

%% file: icassp/setup.tex
\vspace{-2mm}
\section{Experiments}\label{sec:exp}
\vspace{-2mm}
In this section, we demonstrate the efficiency and effectiveness of Omni-sparsity DNN.
With one training process, our supernet simultaneously discovers a set of Pareto models with diversified encoder sparsity ranging from 50\% to 80\%, while keeping the GPU-hours similar to that of a single network.
\vspace{-2mm}
\subsection{Experimental Setup}
\vspace{-2mm}
{\bf Data:} We train our models with the LibriSpeech 960h corpus~\cite{panayotov2015librispeech}. We extract 80-dimensional log Mel-filterbank features from per 25ms window of audio, and strides the window forward in increments of 10ms. We further augment the input features with speed perturbation~\cite{ko2015audio}, at ratio 0.9, 1.0 and 1.1. Spectrum data augmentation~\cite{park2019specaugment} is then added to the features with mask parameter F=27, and 2 time masks with max time-mask ratio p=0.2.
We use the 10.7h Librispeech dev-clean and dev-other data (without augmentation) as the validation dataset for evolutionary search. All the models are trained for 180 epochs unless otherwise specified.\\
{\bf Network Architecture:} We train speech recognizers with the recurrent neural network transducers (RNN-T) models~\cite{graves2012sequence, gao2021extremely}. 
A typical RNN-T consists of an encoder, a predictor and a joiner network. We refer readers to~\cite{he2019streaming} for a detailed explanation on streaming RNN-T models.
In particular, we use the Emformer-based~\cite{shi2021emformer} RNN-T model.
We sandwich 20 layers of Emformer, each with 8 attention heads, 512 hidden units, and 2048 feed forward network dimensions, between two linear projection layers. 
Since the encoder occupies majority of the parameters and computation, we focus only on pruning the encoder. Our model has 77M parameters in total, and the model weights undergo 8-bit post-training quantization during inference.

\vspace{-2mm}
\subsection{Omni-sparsity DNN search space}
\vspace{-2mm}
We construct a search space to allow each sub-network sample a sparsity ratio from $\{0.0, 0.5, 0.6, 0.7, 0.8\}$ at a per layer granularity.
Zero sparsity is not used for sampling random networks or the evolutionary search but only for training the dense network.
To get the sparse layer, we use Adam-pruning criterion and always recompute the pruning mask at each pruning step.
Additionally, we use 8x1 block-wise pruning patterns~\cite{narang2017block} to ensure fast inference on edge devices. 
Empirically, training extremely sparse networks from scratch tends to diverge, we further propose to progressively increase our search space. Specifically, we follow the cubic schedule in~\cite{Zhu2018Prune} with interval 256 (for 256 steps) to dynamic adjust the maximum possible pruning ratio at each step.

%% file: icassp/exp.tex
\section{Results and Discussions}\label{sec:results}
\vspace{-2mm}
\subsection{Comparison with Individually Pruned Models}
\vspace{-2mm}
The Omni-sparsity DNN allows us to find sub-networks of various model sizes with one training job. It presents huge GPU-hour resource saving when compared to independently trained sparse models. In Table~\ref{tab:compare}, we pick four sparsity settings, 50\% to 80\% sparsity ratios, and outline such WER and training time comparisons. Since our supernet is trained with in-place knowledge distillation, we also provide strong baselines of individually trained models with distillation, wherein the teacher is a pre-trained dense model.

As shown in Table~\ref{tab:compare}, the overall training cost of our method is about 3x smaller compared to the training scratch strategy; the models trained from our supernet in achieves the best WER on the test-other dataset. Besides the sparse models listed Table~\ref{tab:compare}, our approach can also flexibly generate other sparse models at no additional training costs, as shown in Figure~\ref{fig:train_time}(a), where we show the sparsity Pareto front.

%

\begin{table}[t]
\centering
\begin{tabular}{l@{ }|c@{ }|c@{ }|c@{ }|c@{ }|c@{ }|c@{ }| c@{ }} 
 \hline
   & \multicolumn{2}{c@{ }|}{Single noKD} & \multicolumn{2}{c@{ }|}{Ours} & \multicolumn{2}{c@{ }|}{Single w/KD} &  \\ 
 Sprsty & \multicolumn{2}{c@{ }|}{WER} & \multicolumn{2}{c@{ }|}{WER} & \multicolumn{2}{c@{ }|}{WER} & size \\ 
\%zeros & clean & other&clean &  other & clean &  other & MB \\\hline
0\% & {\bf 3.4} & 9.3  &  3.7  & {\bf 9.2} & n/a & n/a  & 77 \\\hline
 50\% & 3.7 & 9.7 & 3.9 & {\bf 9.5} &  {\bf 3.6} & 9.8 & 45   \\\hline
 60\% &  3.9 & 10.0 &  3.9 & {\bf 9.6} & {\bf 3.7} & 10.0  & 39  \\\hline
 70\% &  3.9  & 10.6 & 4.0 & {\bf 9.9}  &  {\bf 3.8} & 10.2 & 32  \\\hline
 80\% &  4.2 &  10.9 &  4.1 & {\bf 10.3} & {\bf 3.9} & {\bf 10.3} & 26  \\\hline 
GPU-hr& \multicolumn{2}{c@{ }|}{6912} & \multicolumn{2}{c@{ }|}{\bf 2304} & \multicolumn{2}{c@{ }|}{8448} &\\
 \hline
\end{tabular}
\vspace{-2mm}
\caption{WER of sparse models on Librispeech test sets from individually pruned models with no KL-D, our Omni-sparsity DNN, and KL-D teacher-student trained individual models. Our method generates all models from one supernet.}\label{tab:compare}
\vspace{-2mm}
\end{table}

\vspace{-2mm}
\subsection{Improvement on Training Efficiency}
\vspace{-2mm}
As introduced in Section~\ref{sec:intro}, Wu et al.~\cite{wu2021dynamic} developed DSNN that also jointly optimize four sub-networks that share the underlying weights with sparsity. Due to the uniform layerwise sparsity set-up, DSNN lacks the ability to generate model with new sparsity targets: if DSNN is trained with {0.0, 0.5, 0.7} sparsity ratios, it can not generate a model of 0.55 sparsity. We also compare our Omni-sparsity DNN with DSNN in Figure~\ref{fig:train_time}(b). The dot size corresponds to the model size in Figure~\ref{fig:train_time}(b) -- the largest model has the biggest dot and vice versa. 
We train our supernet for 120 and 180 epochs, respectively. Since DSNN requires training all sub-networks at each training step, it is relatively  computationally expensive. We set the number of epochs of DSNN to be 40, 60, 120 and 180 respectively to show how WER goes with longer training time. 
As Figure~\ref{fig:train_time} shows, our Omni-sparsity supernet converges much faster and yields significantly better WER compared to DSNN with similar amount of training budget.


\vspace{-2mm}
\subsection{Ablation Study}\label{sec:ablation}
\vspace{-2mm}
In Section~\ref{sec:supernettraining}, we introduce three techniques, \emph{in-batch sandwich sampling}, \emph{Adam-pruning} and \emph{adaptive dropout}. In this section, we conduct ablation studies to further verify the effectiveness of each of these training technique. In Table~\ref{tab:finetune_retrain}, we show the supernet is already sufficiently trained with our proposed sandwich sampling, as further training and fine-tuning cannot improve the model WER. In Table~\ref{tab:ablation}, we show that both Adam-pruning and adaptive dropout non-trivially improve the WERs of sparse sub-networks in the supernet. 
\begin{table}[t]
\vspace{-2mm}
\setlength{\tabcolsep}{4pt}
\centering
\begin{tabular}{ c|c|c|c } 
 \hline
 sparsity  & {\bf w/o}  & {\bf w/} Model  & {\bf w/} Supernet   \\
  \% & Finetuning & Finetuning &  Finetuning \\
 \hline
 60\% & 3.9 / 9.6 & 3.9 / 9.7 & 3.9 / 9.6    \\
 \hline
 70\%  & 4.0 / 9.9 & 4.0 / 9.8 & 3.9 / 9.9    \\
 \hline
\end{tabular}
\vspace{-3mm}
\caption{On the effectiveness of efficient sandwich sampling.}
\label{tab:finetune_retrain}
\begin{tabular}{ l|c|c|c|c }
 \hline
 model  & sparsity & \multicolumn{2}{c|}{WER} &  size \\ 
     &  \%      & test-clean & test-other & MB \\\hline
 Baseline &  & 4.2 & 10.7  &  \\ 
 \ + Adam-pruning & 60\% & 4.0 & 10.0 & 39   \\
 \ \ + Adaptive Dropout & & 3.9 & 9.6  &  \\
 \hline
  Baseline &  & 4.4 & 11.1  &  \\
  \ + Adam-pruning & 70\% & 4.1 & 10.4 & 32   \\
  \ \ + Adaptive Dropout &  & 4.0 & 9.9  &  \\
 \hline
\end{tabular}
\vspace{-3mm}
\caption{On the impact of adam prune and adaptive dropout.}
\label{tab:ablation}
\end{table}



%% file: icassp/conclusion.tex
\label{sec:ablation}
\vspace{-2mm}
\section{Conclusion}
\vspace{-2mm}
In this work, we propose the Omni-sparsity DNN, which incurs similar GPU-hours of training time as one single sparse model does, and yet generates various optimized sparse models to fit the constraints of a variety of edge devices. We proposed several effective methods to train the Omni-sparsity DNN, balancing training speed and model stability with model accuracy. These include Adam-pruning, adaptive dropouts, and in-batch sandwich sampling. We outline how evolutionary search can be efficiently used to find optimal sub-networks in the Omni-sparsity DNN. 